# Three-dimensional structure of a single colloidal crystal grain studied by coherent x-ray diffraction

J.Gulden[1], O.M. Yefanov[1], A.P. Mancuso[1+], R. Dronyak[1], A. Singer[1], V. Bernátová[1,2], A. Burkhardt[1], O. Polozhentsev[1,3], A. Soldatov[3], M. Sprung[1], and I.A. Vartanyants[1,4*]

[1]*Deutsches Elektronen-Synchrotron DESY, Notkestraße 85, D-22607 Hamburg, Germany;*
[2]*Faculty of Science, University of Pavol Jozef Šafárik, Šrobárova 2, 041 80 Košice, Slovak Republic;*
[3]*Southern Federal University, Zorge str. 5, 344090 Rostov-Don, Russia;*
[4]*National Research Nuclear University, "MEPhI", 115409 Moscow, Russia*

Friday, 24 June 2011

**ABSTRACT**

A coherent x-ray diffraction experiment was performed on an isolated colloidal crystal grain at the coherence beamline P10 at PETRA III. Using azimuthal rotation scans the three-dimensional (3D) scattered intensity in reciprocal space from the sample was measured. It includes several Bragg peaks as well as the coherent interference around these peaks. The analysis of the scattered intensity reveals the presence of a plane defect in a single grain of the colloidal sample. We confirm these findings by model simulations. In these simulations we also analyze the experimental conditions to phase 3D diffraction pattern from a single colloidal grain. This approach has the potential to produce a high resolution image of the sample revealing its inner structure, with possible structural defects.

---

[+]Present address: European XFEL GmbH, Albert-Einstein-Ring 19, 22761, Hamburg, Germany

[*]Corresponding author: ivan.vartaniants@desy.de

# I. Introduction

Direct methods in crystallography (Giacovazzo *et al.*, 2002) have revolutionized the field of structure determination in protein crystallography. Based on the simultaneous measurement of Bragg peaks, while rotating the crystal and powerful computational phasing methods, the structure of proteins to sub-Angstrom resolution can be revealed. Crystallographic methods for structure determination are based on the assumption of perfect crystallinity of the sample. Unfortunately, structural defects, in the form of vacancies, dislocations, and stacking faults – which are usually present in protein crystals – are additional complications which lower the fidelity of structure determination. From that respect understanding of the local structure of protein crystals on a length scale below one micron is an important task in structural biology.

More then half a century ago Sayre (Sayre, 1952) proposed to phase crystallographic data by measuring information between the Bragg peaks in reciprocal space. This method was later termed an oversampling method contrary to conventional sampling at Bragg peaks in crystallography. The first experimental demonstration (Miao *et al.*, 1999) showed its potential to reveal the structure of a non-periodic, finite size sample illuminated by a coherent radiation by applying iterative phase retrieval algorithms (Fienup, 1982; Elser, 2003). Coherent x-ray diffraction imaging (CXDI) was applied later to a variety of organic and non-organic samples (see for reviews (Nugent 2010; Mancuso *et al.*, 2010)) including crystalline particles (Williams *et al.*, 2003; Pfeiffer *et al.*, 2006; Robinson & Harder, 2009).

Measuring diffraction patterns from a coherently illuminated finite size crystalline sample locally around a selected Bragg peak gives information about the shape and a uniform distribution of strain in this particle (Pfeiffer *et al.*, 2006; Robinson & Harder, 2009). This method was recently extended to image a finite region of a two-dimensional (2D) colloidal crystal (Gulden *et al.*, 2010). In this proof-of-principle experiment it was demonstrated that inversion of the coherent diffraction pattern containing several Bragg peaks gives the position of individual scatterers (colloidal spheres in this case) as well as the core structure of the defects in the sample. Here we extend these findings to a single 3D grain of a colloidal crystal illuminated by coherent x-rays. We present the first results of the experimental realization of Sayre's original idea of collecting 3D crystallographic data and measuring the scattered intensity not only at the positions of the Bragg peaks but also in the region of reciprocal space between the Bragg peaks. Reconstruction of these data can provide 3D information about the actual positions of the scatterers in the colloidal crystals including possible structural defects.

High resolution 3D imaging of internal structure of mesoscopic materials, such as colloidal crystals, by x-ray methods is a challenging problem (Bosak *et al.*, 2010; van Schooneveld *et al.*, 2011). Development of new imaging approaches based on coherent x-ray scattering methods is quite important for understanding the internal structure of these novel materials. Self-organized colloidal crystals can be used as the basis for new functional materials such as photonic crystals, which may find applications in future solar cells, LEDs, lasers or even as the basis for circuits in optical computing and communication. For these applications, crystal quality is crucial and monitoring the defect structure of these real colloidal crystals is essential (Verhaegh *et al.*, 1995; Blanco *et al.*, 2000; Smajic *et al.*, 2000; Vlasov *et al.*, 2000; Yannopapas *et al.*, 2001; Rengarajan *et al.*, 2005; Hilhorst *et al.*, 2009).

The paper is organized as follows. In the next section we will describe the experiment performed at the synchrotron source PETRA III, in section three we discuss the outcome of that experiment. In section four we present results from simulations performed on a model system of a colloidal grain. The fifth section presents a reconstruction from the simulated data. This is followed by the conclusion and outlook.

**II. Experiment**

The coherent x-ray scattering experiment was performed at the Coherence Beamline P10[1] of the PETRA III facility at DESY in Hamburg during commissioning. The incident energy of the x-rays was chosen to be 7.9 keV to have a high penetration depth through the colloidal crystal under investigation and to map several Bragg peaks onto a single 2D detector.

Silica microspheres were synthesized and coated with 1-octadecanol to provide steric stabilization as described by (Verhaegh & van Blaaderen 1994). The particle diameter of about 230 nm was determined by small-angle x-ray scattering in a dilute suspension using the separation between the form-factor minima (Petukhov *et al.*, 2004) and for the sedimentary colloidal crystals from the position of the Bragg peaks (Petukhov *et al.*, 2002). The particle size polydispersity of 4.1% was determined by transmission electron microscopy. To grow colloidal crystals, the microspheres were dispersed in cyclohexane, placed in a vial and kept vertically for sedimentation and self-organisation. The solvent slowly evaporated on the time scale of several months. The top layer of the dry sediment (of about 5 mm thickness) showed characteristic optical Bragg reflections under white light illumination. A piece of this dry sediment was mechanically crushed to yield a powder of small grains.

Individual grains of such colloidal crystals were picked up manually, using a micromanipulator (PatchMan NP2). They were attached by an adhesive to the tip of a 10 μm diameter carbon fiber, which was mounted on the sample holder. Afterwards they were characterized using an optical microscope (Leica M165C). Several grains, with varying sizes from 3 to 10 μm in diameter, were prepared using this method for characterization by coherent x-ray scattering.

---

[1] Information about Coherence Beamline at PETRA III can be found at:
http://hasylab.desy.de/facilities/petra_iii/beamlines/p10_coherence_applications/index_eng.html

The scattering geometry of the experiment is shown in Fig. 1. It includes a pair of coherence defining and guard slits located at distance 80.6 cm and 29 cm in front of the colloidal sample. The detector was positioned 5.3 m behind the sample in the far-field. The sample was mounted on a goniometer, which allows rotation of the sample around the vertical axis. An evacuated tube was positioned between the sample and the detector to reduce air scattering. The diffraction data were recorded using a charge-coupled device (CCD) detector (Roper Scientific, PI-LCX) with a pixel size of 20×20 µm$^2$ and 1300×1340 pixels with a total field of view of 26×26.8 mm$^2$. The maximum detector resolution in this geometry was $q_{max}$ = 98 µm$^{-1}$. To protect detector from the direct beam, a round beamstop of 2.5 mm in diameter was used in front of the detector.

Several individual grains of colloidal crystal were measured in azimuthal angular scans (see Fig. 1). In total 180 diffraction patterns were obtained for each sample, in one degree increments covering the entire reciprocal space. For each angular position from 10 to 100 diffraction patterns were measured to increase the dynamic range and improve the signal to noise ratio. The exposure times ranged from 0.2 s up to 30 s per image depending on the scattered intensity. Measured diffraction patterns at each angular position were averaged and corresponding dark field images were subtracted. In addition, to obtain detailed information, fine scans with a step size of 0.2° were made around selected Bragg peaks.

**III. Results**

Typical diffraction patterns measured at different relative azimuthal angles of rotation $\Delta\theta$ from two different colloidal crystal grains are shown in Fig. 2. Several Bragg peaks are clearly visible in these diffraction patterns. The most pronounced ones are the hexagonal set of (220) Bragg peaks typical for the scattering on an fcc structure. Due to the finite size and internal structure of the crystal grains, each of the measured Bragg peaks breaks into a complicated speckle pattern (see inset in Fig. 2*c*). This is similar to observations in our previous experiments on coherent scattering from finite crystals (Williams *et al.*, 2003; Pfeiffer *et al.*, 2006; Gulden *et al.*, 2010; Mancuso *et al.*, 2009). In addition to Bragg peaks, we observed strong speckles over the entire diffraction pattern, which are typical for coherent scattering experiments on colloidal samples (Xinhui Lu *et al.*, 2008). In our experiment, we attribute this to a contribution from the disordered part of the sample that may be present in the grain.

A strong $q$-dependence of the diffraction pattern, due to the variation of the form-factor from the individual colloidal spheres is observed in Fig. 2 as well. Due to the finite size distribution of these spheres the $q$-dependence of the form-factor never reaches zero. A strong enhancement of the speckle pattern in the range of the zero-order of the form-factor compared to the higher $q$-values is clearly seen in Fig. 2. The strong flares in each diffraction pattern appear due to parasitic scattering from the slits located in front of the sample and scattering from the carbon fibers. Since the carbon fibers were not perfectly oriented vertically in the beam, the scattering produced by them changed with the variation of the azimuthal angle $\Delta\theta$.

The diffraction patterns produced by scattering of a coherent x-ray beam from two different colloidal grains look similar (see Fig. 2). However, there was one major difference in our observations. For one grain, we saw strong streaks through the Bragg peaks at a relative rotation angle of $\Delta\theta = 80°$ (see Fig. 2*b*), which are similar to our earlier findings for a 2D film of colloidal crystal (Gulden *et al.*, 2010). At the same time, no streaks were observed in the diffraction patterns during the entire scan for the other grain (see Figs. 2*c* and 2*d*). Such streaks are an indication of a possible defect in the crystalline sample (Hilhorst *et al.*, 2009). From this result, the conclusion may be made that the first grain contains structural defects and the second one does not. However, such a conclusion about the structure of the sample deduced only from the analysis of the 2D diffraction patterns can be misleading, as explained later.

We combined 2D diffraction patterns measured from the second grain into a 3D data set using nearest-neighbour interpolation. Fig. 3*a* shows a representation of the measured data in 3D, showing three orthogonal planes (cuts) through reciprocal space on a logarithmic color scale, as well as an isosurface of the scattered intensity. To our surprise, in this 3D representation of the coherent scattering from the second grain, the streaks became visible indicating the presence of a defect in this sample as well. By analyzing the structure of the coherently scattered intensity we can even make conclusions about the type of defect present. The additional features in the 3D diffraction pattern have the shape of thin rods passing through the (111) Bragg peaks and this is a clear indication that the colloidal grains contain a plane defect in the [111] plane. This can not be a linear defect, for example a linear dislocation, because that will produce additional scattering in reciprocal space in the form of a plane (see, for example (Vartanyants *et al.*, 2008), where crystal truncation planes were observed in reciprocal space due to scattering on the line edges of the sample). Stacking faults (Warren, 1990) are typical plane defects known for fcc structures. We attribute these additional rods observed in the 3D intensity distribution to scattering on a stacking fault present in our grain. In order to confirm these observations, we performed simulations based on a model of a colloidal sample similar to that used in the experiment and present the results of these simulations in the following sections.

In Fig. 3*c* the 3D structure of the scattered intensity in the vicinity of a 220 Bragg peak for the same sample obtained as a result of a fine angular scan is presented. A complicated 3D intensity distribution can be observed in this figure. This implies that the 3D structure of the colloidal grain under investigation is also complicated.

Attempts to reconstruct the full 3D data set shown in Fig. 3*a* were, unfortunately, unsuccessful. We attribute this to the large amount of the missing data covered by the beamstop, which is known to limit the reconstruction process (Thibault *et al.*, 2006), as well as to the parasitic scattering from the slits and the carbon fiber. In the following sections we discuss the conditions that are necessary for a successful reconstruction of a single colloidal grain, based on our model simulations.

**IV. Model**

Diffraction patterns of coherent x-ray scattering at 7.9 keV incident photon energy for a model colloidal sample were simulated. The model was composed from colloidal spheres made from silicate with a diameter of 230 nm arranged in an fcc crystal lattice. The shape of the whole grain was taken in the form of an ellipsoidal particle with a total size of 5x6x7 μm$^3$ (see inset in Fig. 4*a*), which is similar in size to the grains used in the experiment. In addition, a single defect in the form of a stacking fault was introduced inside our model grain particle.

We simulated 2D diffraction patterns from this structure for different angles of azimuthal rotation Δθ from 0° to 180° with an angular increment of 0.25°. Diffraction patterns were calculated assuming kinematical scattering and plane wave illumination of the sample. All diffraction patterns were combined into a 3D data set using the same approach as for the experimental data. Fig. 3*b* shows the scattered intensity in 3D reciprocal space as one in Fig. 3*a* for experimental data. Several common features can be identified while comparing these two images. The most prominent is the hexagonal set of the (220) Bragg peaks. Around each Bragg peak we see strong fringes due to coherent scattering from the finite size crystalline particle (see magnified view in Fig. 3*d*). In addition, six parallel streaks going through the set of (111) Bragg peaks can be observed, which originate from the scattering from the stacking fault introduced in our model. The structure and position of these streaks in reciprocal space is the same as in the experimental data (see Fig. 3*a*). The same features are well distinguished in a horizontal ($q_x$-$q_y$) cut through 3D reciprocal space obtained in the model simulations (see Fig. 4*a*). This gives us confidence that we have indeed observed a defect in the form of a stacking fault in a single grain of a colloidal particle in our experiment.

## V. Reconstruction

In order to determine the necessary experimental procedure to image the structure of a single colloidal crystal grain in coherent diffraction experiment the simulated 3D scattered intensities shown in Fig. 3*b* were phased using an iterative phase retrieval algorithm.

Starting from a rectangular support, we applied the shrink-wrap (SW) algorithm (Marchesini *et al.*, 2003) to find more realistic estimate of the shape of the object. For each reconstruction 100 iterations of hybrid input-output (HIO) algorithm (Fienup, 1982) were followed by 50 iterations of the error-reduction (ER) algorithm. After each $900^{th}$ iteration the SW algorithm was applied again to optimize the support. Figs. 4*b*, 4*c*, and 4*d* show the result of our reconstruction, after a total of 2700 iterations. In the reconstruction no significant difference to the initial model could be found. An outer isosurface of the grain obtained from the reconstruction (Fig. 4*b*) is almost identical to the initial model (see inset in Fig. 4*a*). Clearly, the crystallinity of the sample was reconstructed, showing the positions of each individual scatterer, which can be seen in a cut through the reconstructed sample in Figs. 4*c* and 4*d*. Even the defect, in the form of a stacking fault as a break in the "correct" ABC ordering is resolved, in these ideal conditions.

In our experiment the angular step size was one degree, which is four times bigger than in our simulations. This sparse angular sampling as well as the necessary beamstop may prevent successful convergence of the algorithm (Huang *et al.*, 2010). To investigate these complications in more detail, we performed another set of simulations where we addressed these questions of missing data and sparse angular sampling on the outcome of the phase retrieval.

We took a subset of the simulated data with 180 diffraction planes in total with one degree rotational increment similar to the conditions of our experiment. The central part of the diffraction pattern was not used in the reconstruction to simulate a beamstop that covers the central speckle and two additional fringes. Fig. 5*a* shows a ($q_x$-$q_y$) cut through the 3D diffraction pattern where artefacts appear due to the sparse angular sampling. Comparison with the previous simulation, when a finer step size of 0.25° was used (see Fig. 4*a*) clearly shows larger regions of missing data at high *q*-values.

To address the problem of missing data in the low *q*-region as well as to obtain an estimate of the shape of the crystalline grain we used as the initial step for the 3D reconstruction the scattered intensity around a single Bragg peak (see inset in Fig. 5*a*). As we demonstrated in our experiment, this region of reciprocal space in the vicinity of the Bragg peak can be measured with fine angular steps and without a beamstop. We applied the same reconstruction procedure as described before to these data and obtained a rough estimate of the support for a colloidal grain. We should note here, that this procedure is clearly limited to crystalline particles. For example, if crystalline particle is surrounded by a non-crystalline part, only the shape of the crystalline part will be obtained by applying this approach. The found support was used to evaluate the intensity in the central part of the diffraction pattern with missing data. We substituted this missing central part of the diffraction pattern by the properly scaled central speckle obtained from the support (Nishino *et al.*, 2008) and allowed to free evolve the intensities in all other non defined pixels. The reconstruction was based on the repetition of 100 and 50 iterations of the HIO and ER algorithms, respectively.

In Figs. 5*b*, 5*c* and 5*d* the results of the reconstruction after 2700 iterations is presented. The crystallinity of the sample and, importantly, the stacking fault are still clearly resolved. In comparison with the result presented in the Fig. 4, the overall image quality become worse, particularly a non-uniform density variation across the sample is noticeable. These artefacts appear due to unconstrained modes in the central part of the diffraction pattern (Thibault *et al.*, 2006). Our simulations have shown as well that attempts to reconstruct the structure of the colloidal grain with a larger region of missing data in the central part of the diffraction pattern were not successful. This indicates that in order to obtain reliable reconstruction in the conditions of our experiment the beamstop should not be substantially bigger than the central speckle.

The performed simulations were made with noise free data. Our experience shows that a few per cent of noise do not change the results significantly.

**VI. Summary and outlook**

In summary, coherent x-ray diffraction patterns from isolated colloidal crystal grains were recorded at the coherent x-ray beamline at PETRA III, covering several Bragg peaks as well as the coherent interference signal between the peaks. We measured different grains in a rotation series. These diffraction patterns were combined into a 3D data set containing the structural information about the object in reciprocal space. We observed streaks in reciprocal space that are clear indications of a plane defect in our grain.

We confirmed our observations by simulations performed on a model of a single grain of a colloidal particle with a defect in the form of a stacking fault. The 3D distribution of intensity from that model has similar features to those observed in our experiment that gives fidelity to our conclusions. We inverted this intensity distribution using iterative phase retrieval methods. The results of our 3D reconstruction reveal the shape and the inner structure of the grain. The position and the structure of a stacking fault defect were obtained with high resolution as well.

We propose to use the results of the reconstruction of the data collected around the Bragg peak in order to determine the tight support which is required to treat a missing data problem. Our simulations show that diffraction patterns collected in the conditions close to our experiment, with the rotation step of one degree and the beamstop covering the central speckle and two additional fringes are sufficient for a successful reconstruction. With this knowledge, in future experiments, we plan to collect a set of data that will be sufficient to retrieve the structure of individual grains, showing the positions of the individual colloidal particles including possible structural defects.

**Acknowledgments**

J. Hilhorst and A. Petukhov are gratefully acknowledged for providing colloidal crystal samples and useful discussions. The authors would like to thank M. Dommach and S. Bondarenko for the technical support and their help during the experiment. We would like also to thank E. Weckert for the support and interest during the work on this project as well as for the use of the MOLTRANS code for simulations. Part of this work was supported by BMBF Proposal 05K10CHG "Coherent Diffraction Imaging and Scattering of Ultrashort Coherent Pulses with Matter" in the framework of the German-Russian collaboration "Development and Use of Accelerator-Based Photon Sources".

# Figure Captions

**Figure 1.** (Color online) Schematic view of the CXDI experiment showing the guard slits, the colloidal sample and the detector.

**Figure 2.** (Color online) Selected diffraction patterns measured at different relative angles $\Delta\theta$ of the azimuthal rotation for the colloidal grain one (a, b) and the colloidal grain two (c, d). The inset in panel (c) presents a magnified view of the region around the Bragg peak shown by the white square.

**Figure 3.** (Color online) (a) A 3D representation of the measured scattered intensity showing three orthogonal planes and an isosurface, (b) the same view for the simulated scattered intensities from the model. The length of the arrows corresponds to 30 $\mu m^{-1}$. (c) An isosurface of measured scattered intensities around a 220 Bragg peak. (d) An isosurface of simulated scattered intensities around a 220 Bragg peak. The length of the arrows corresponds to 2.5 $\mu m^{-1}$.

**Figure 4.** (Color online) (a) The ($q_x$ - $q_y$) cut in reciprocal space showing the distribution of the scattered intensity on a logarithmic scale from the model colloidal sample presented in the inset. (b) An isosurface of the reconstructed object, the length of the arrows corresponds to 2 $\mu m$. (c, d) The orthogonal cuts through the reconstructed object. The stacking fault is clearly seen in (d).

**Figure 5.** (Color online) (a) The ($q_x$ - $q_y$) cut in reciprocal space obtained from simulated 2D diffraction patterns with one degree increment. The missing region covers the central speckle and two first fringes. (inset) A slice through 3D scattered intensity around a 220 Bragg peak. (b) An isosurface of the reconstructed object, the length of the arrows corresponds to 2 $\mu m$. (c, d) The orthogonal cuts through the reconstructed object.

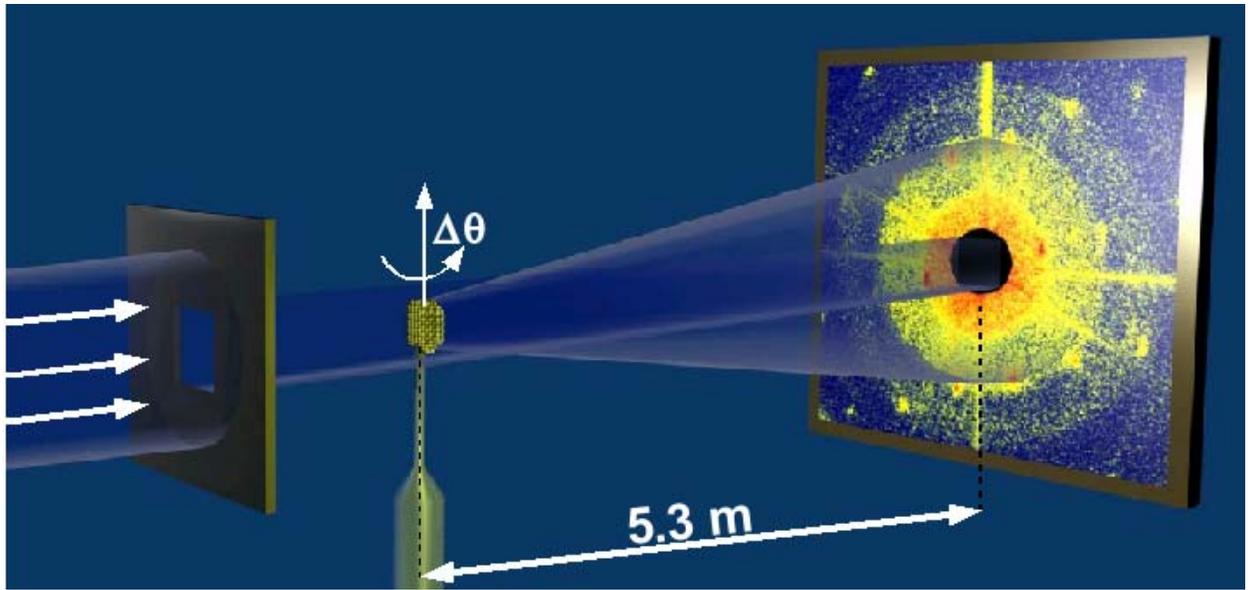

**Figure 1**

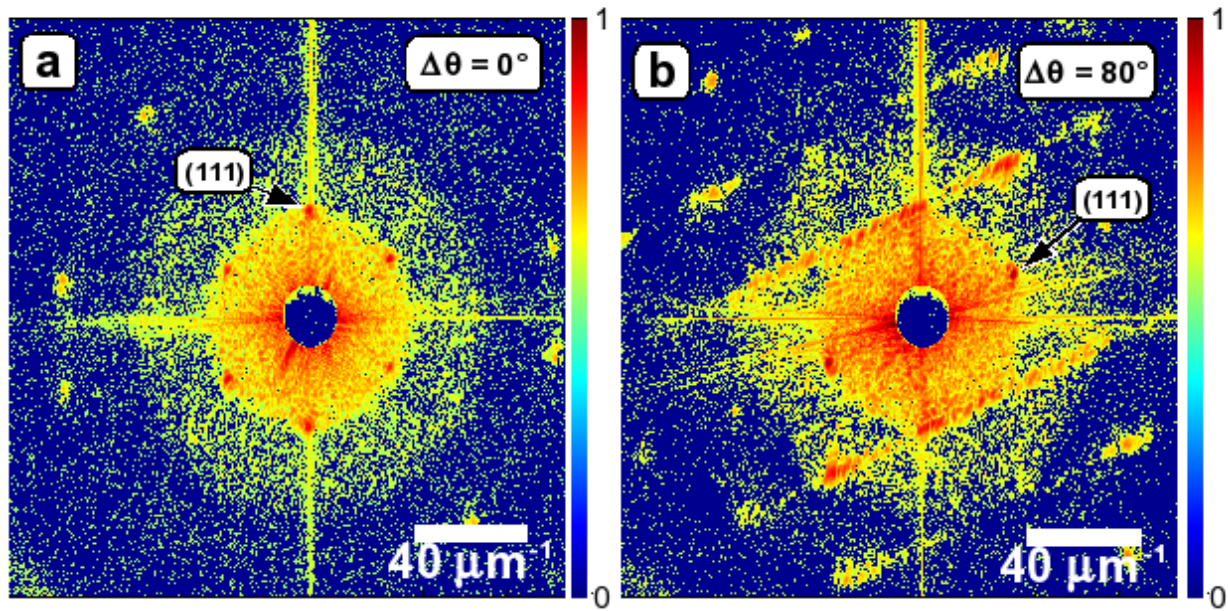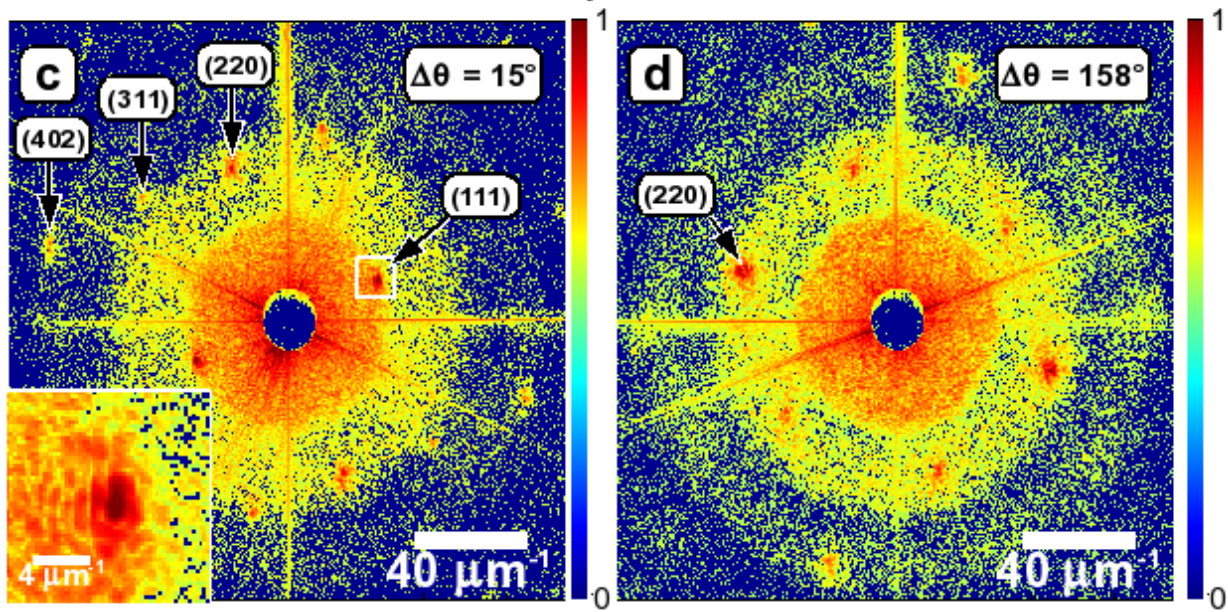

**Figure 2**

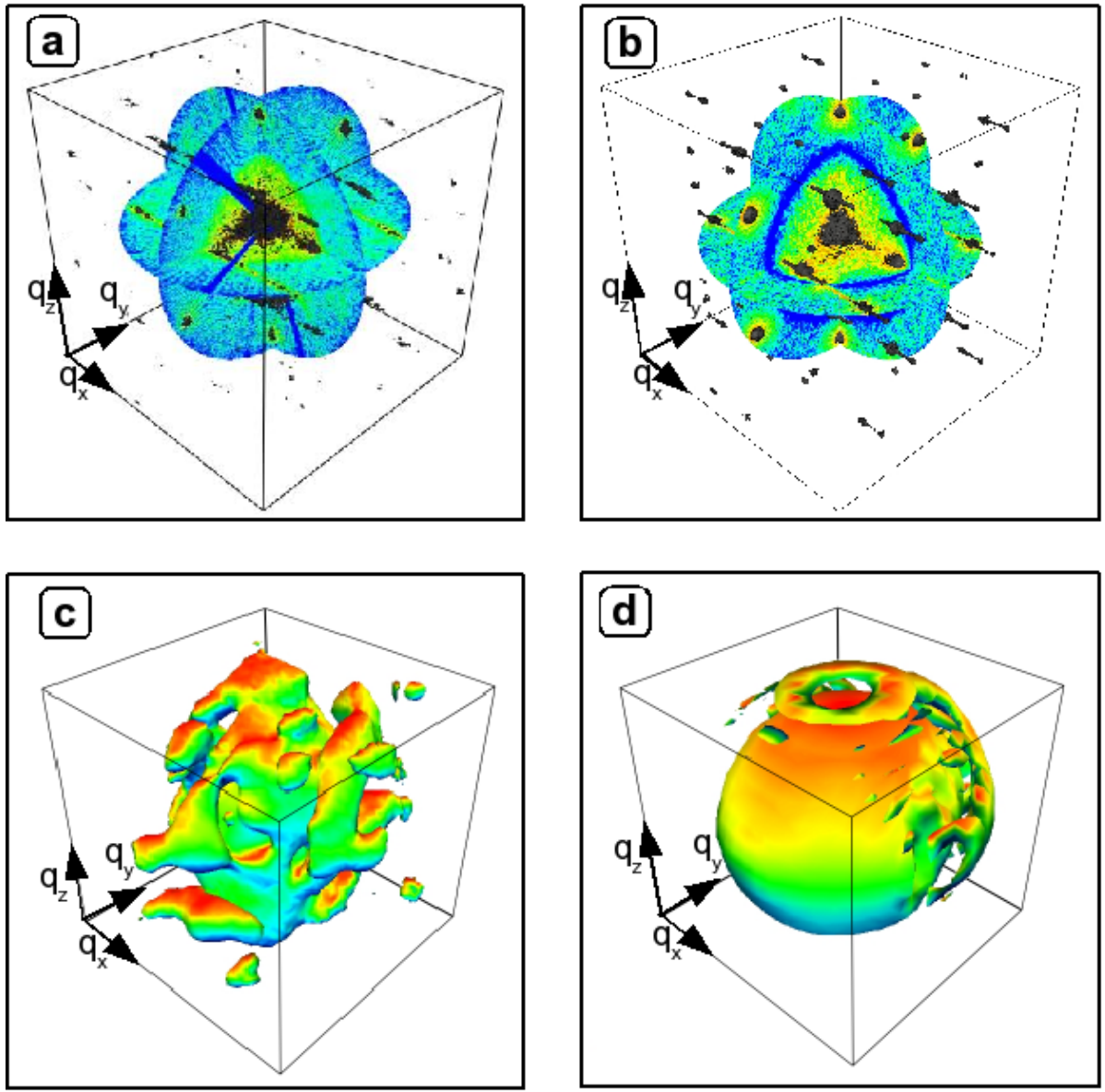

**Figure 3**

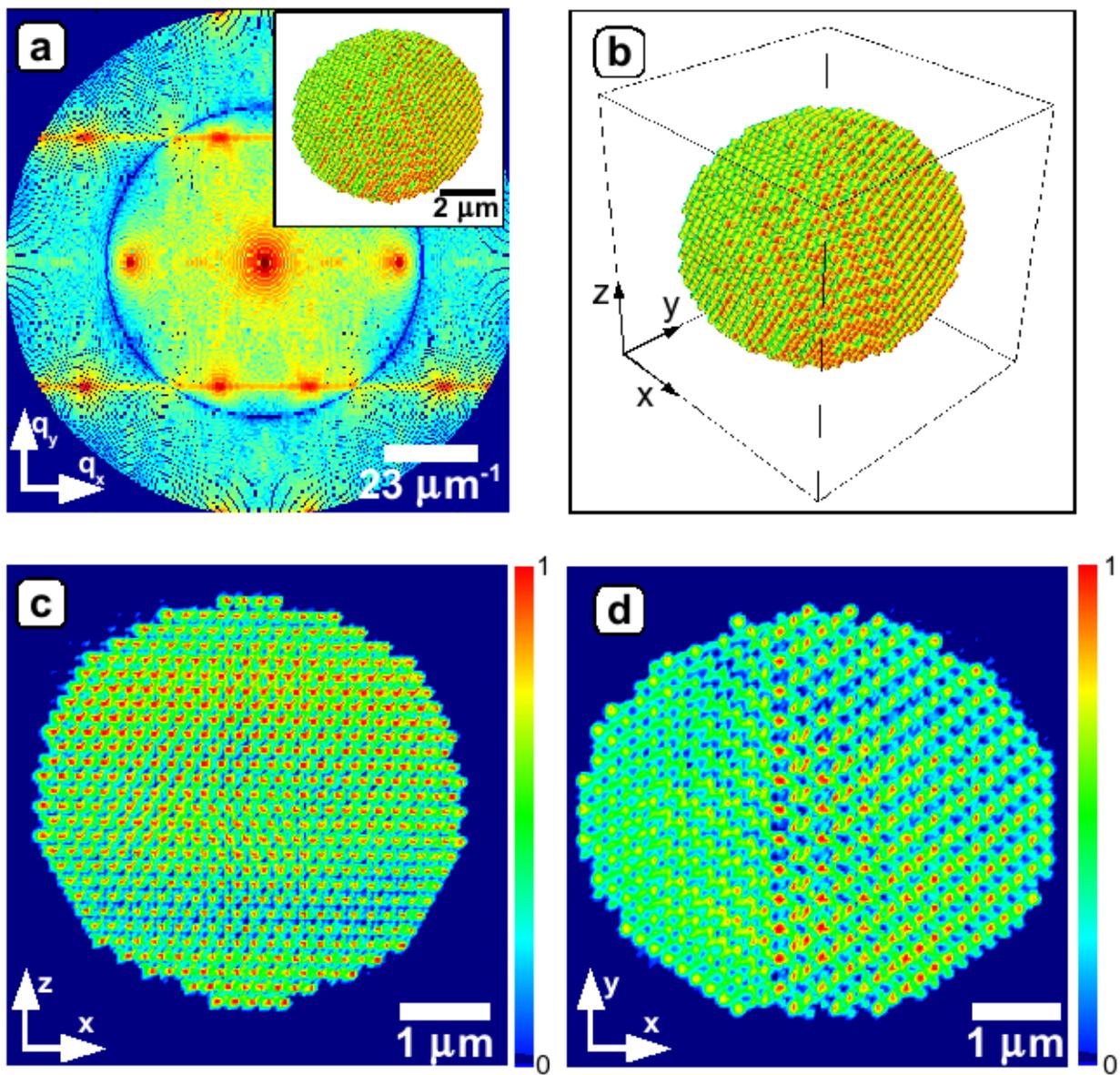

**Figure 4**

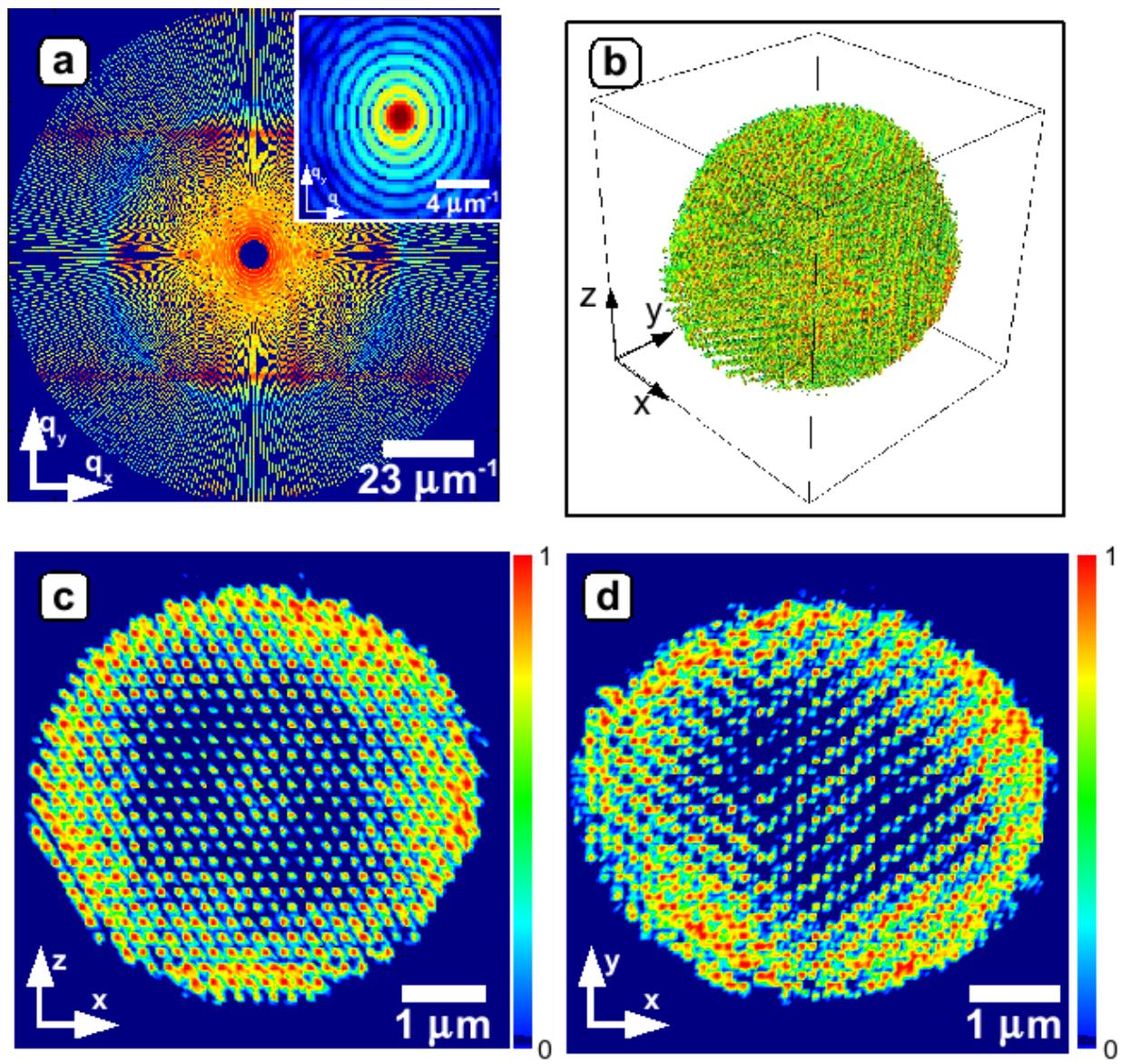

**Figure 5**